\def\year{2020}\relax
\documentclass[letterpaper]{article} 
\usepackage{aaai20}  
\usepackage{times}  
\usepackage{helvet} 
\usepackage{courier}  
\usepackage[hyphens]{url}  
\usepackage{graphicx} 
\def\UrlFont{\rm}  
\usepackage{graphicx}  
\usepackage{color}

\newcommand{\eg}{{e.g.,} }
\newcommand{\ie}{{i.e.,} }

\usepackage{subcaption}
\usepackage[normalem]{ulem}

\usepackage{enumerate}

\usepackage[hyphens]{url} 

\title{Web Routineness and Limits of Predictability: \\Investigating Demographic and Behavioral Differences Using Web Tracking Data}

 \author{Juhi Kulshrestha\textsuperscript{\rm 1}, Marcos Oliveira\textsuperscript{\rm 1}, Orkut Karaçalık\textsuperscript{{\rm 1},{\rm 2}}, Denis Bonnay\textsuperscript{\rm 3}, Claudia Wagner\textsuperscript{{\rm 1},{\rm 2}} \\
 \textsuperscript{\rm 1}GESIS - Leibniz Institute for the Social Sciences, Germany\\
 \textsuperscript{\rm 2}University of
 Koblenz-Landau, Germany\\
 \textsuperscript{\rm 3}Université Paris Nanterre, France\\
 }

\usepackage{xcolor}
\usepackage{soul}
\definecolor{reddish}{HTML}{FBB4AE}
\definecolor{blueish}{HTML}{B3CDE3}
\definecolor{magentish}{HTML}{FF00AA}
\definecolor{greenish}{HTML}{a1d99b}
\definecolor{yellowish}{HTML}{FFFF99}

\usepackage{amsmath}

\usepackage{booktabs}

\begin{document}

\maketitle

\begin{abstract}

Understanding human activities and movements on the Web is not only important for computational social scientists but can also offer valuable guidance for the design of online systems for recommendations, caching, advertising, and personalization. In this work, we demonstrate that people tend to follow routines on the Web, and these repetitive patterns of web visits increase their browsing behavior's achievable predictability.  We present an information-theoretic framework for measuring the uncertainty and theoretical limits of predictability of human mobility on the Web. We systematically assess the impact of different design decisions on the measurement. We apply the framework to a web tracking dataset of German internet users. Our empirical results highlight that individual's routines on the Web make their browsing behavior predictable to 85\% on average, though the value varies across individuals. We observe that these differences in the users' predictabilities can be explained to some extent by their demographic and behavioral attributes. 

\end{abstract}

\section{Introduction}
The World Wide Web is an immense space containing more than six billion websites that keep popping up and dying every second~\cite{size_of_the_web_2020}. We spend more than a quarter of our day in this fluid space, performing activities such as shopping, reading the news, or interacting with friends~\cite{digital_2019}. Though the Web is an indispensable part of people's lives today, we still do not fully understand the dynamics of people's web browsing behavior. Such understanding is vital to improve the way we navigate the Web and provides grounds for algorithms to assist users in their exploration of the cyberspace. 

In the case of physical space, researchers have already started to understand the fundamental mechanisms governing the dynamics of human mobility~\cite{Gonzalez_2008,Song_2010a}. 
Despite the complexity of our decision-making processes, the way we move in physical space has been shown to present regularities at varying spatiotemporal scales~\cite{Pappalardo2015returners,Alessandretti2018,barbosa2018}. 
Such regularities emerge from our daily routines and constraints, which in turn make our movements highly predictable~\cite{song2010limits}. 
Understanding these characteristics of human mobility is critical for managing crucial aspects of our society, including public health, transportation, urban planning, to name a few. 

With the increasing reliance on the Web, people have switched to performing many everyday activities online rather than offline.
With people living more and more of their lives online, the way they traverse the Web affects how they gather information and interact with others, influencing different aspects of society such as education~\cite{Ravizza2017}, mental health~\cite{Culjak2016}, news consumption~\cite{moller2019explaining,scharkow2020social}, and political participation~\cite{Stier.2020}. 
Uncovering the fundamental properties of web browsing behavior has become imperative to understand society better.

Recently, examining and modeling the dynamics of human movements on the Web have started to receive more attention, with researchers leveraging models developed for mobility in physical space to study them~\cite{zhao2014scaling,zhao2015non,zhao2016unified,barbosa2016returners,hu2018life,hu2019return,hazarie2019uncovering}. 
Previous works have uncovered underlying mechanisms in web mobility, such as preferential return and exploration.
However, much research has predominantly focused on characterizing distributions
and time series.
Researchers have often overlooked heterogeneity in users' characteristics and often limited their analysis to aggregated levels and specific platforms. 
Therefore, the individual-level regularities in web browsing behavior are as yet poorly understood. 


%

In this work, we show that individuals tend to follow routines on the Web---a property we call  \mbox{\textit{web routineness}}. These repetitive patterns of online visitation considerably increase the achievable predictability concerning users' browsing behavior.  We present an information-theoretic framework for estimating the theoretical limits of predictability of people's web mobility based on previous work~\cite{song2010limits}. By examining web tracking data from $2,148$ users, we show that individuals' {web routineness} makes their browsing behavior predictable to 85\% on average. However, we show that the routineness and predictability vary considerably across users and that users' demographic and behavioral differences can explain part of this variation. 



Our work not only has implications for computational social scientists but also computer scientists and industry practitioners. While much prior work~\cite{pirolli1999distributions,narvekar2015predicting,deshpande2004selective,mabroukeh2009semantic,manavoglu2003probabilistic,pitkow1999mininglongestrepeatin,su2000whatnext,awad2007web,khalil2009integrated} has focused on developing specific algorithms for predicting the next visited location on the Web, our work focuses on estimating the theoretical upper limits for correctly predicting an individual's next visited web location based on their browsing trajectories. Our work can inform researchers and practitioners about what performance they can aspire to achieve for which web prediction task by leveraging the routines users follow in their browsing trajectories. 
Moreover, our results also have implications for privacy researchers since the high predictability for most users indicates that attackers can also predict what location the user would visit next, having observed the routines users follow from their past browsing histories. 

Finally, we summarize the  contributions of this paper:

\begin{itemize}
    \item Building upon prior work on mobility in physical space~\cite{song2010limits,Smith2014,ikanovic2017alternative}, we present an adapted information-theoretic framework for estimating the theoretical limits of predictability of humans' mobility on the Web. 
    
    \item We describe how to construct stationary and non-stationary trajectories from users' raw web tracking data.
    
    \item We systematically assess the impact of different design decisions (\eg temporal and spatial resolution, and stationarity in trajectories) on the measurement and provide practical guidelines for applying the framework to users' web tracking data.
    
    \item We investigate users' routineness and predictability on the Web by applying the framework to a web tracking dataset of $2,148$ German internet users. We show that individuals exhibit routineness while browsing the Web and that these routines or repetitive patterns make their browsing behavior predictable to 85\% on average. Our results revealed considerable variability in the routineness and predictability across users. We show that users' demographic and behavioral differences can explain part of this variation. 
    
    \item Finally, we make available our code\footnote{\url{https://github.com/gesiscss/web_tracking}} along with tutorial notebooks and data~\cite{webtrackingdata} to the researcher community to encourage further research on mobility on the Web.
    
\end{itemize}

\section{Background \& Related Work}

\subsubsection{Human mobility in physical space} 
Research on human mobility has shown that it is characterized by a heterogeneity of travel patterns~\cite{Gonzalez_2008,Pappalardo2015returners}, a high degree of predictability~\cite{song2010limits}, 
and a strong tendency of humans' to spend most of their time in a few locations~\cite{Song_2010a,Alessandretti2018}. 
We leverage previous work on human mobility~\cite{song2010limits,ikanovic2017alternative,Smith2014} to present an adapted predictability measurement framework for estimating the upper limits of predictability of people's mobility on the Web.

\subsubsection{Adapting physical space mobility models for cyberspace}
Prior work on investigating mobility in cyberspace, using concepts from physical mobility literature, found several interesting similarities, such as similar scaling properties of return visitation patterns ~\cite{barbosa2016returners,hu2018life,hu2019return}, similar superlinear scaling relation between mean frequency of visit and its fluctuation~\cite{zhao2014scaling}, similar focus on a small, stable number of mobile apps~\cite{de2019apps} and number of familiar locations in physical space~\cite{Alessandretti2018}, and similar memory-based random walk dynamics~\cite{zhao2016unified}.
Several prior studies have applied the framework proposed by Song et al.~\cite{song2010limits} in the online context to study entropy and predictability for mobility in multiplayer online games~\cite{sinatra2014entropy}, mobile phone users' traffic usage~\cite{tao2019behavior}, and textual conversations~\cite{bagrow2019information}, information cascades~\cite{kolli2017quantifying}, and movement of users across communities ~\cite{hu2018life} on specific social media platforms. 
Instead, we focus on the routines that people follow on the web to estimate the theoretical upper limit of predictability of user's next visited location while traversing the whole Web.
We find these cyberspace limits to be lower than the limits computed for physical mobility~\cite{song2010limits,ikanovic2017alternative,zhao2015non}. 
Furthermore, we observe differences in the predictability of users with different demographic and behavioral characteristics, which 
were not observed for mobility in the physical space.

\subsubsection{Web navigation prediction algorithms}
Another complementary line of prior work has focused on developing specific algorithms for predicting the next web access by users, given users' web access sequences. 
For this task, prior work has leveraged the routines in people's browsing trajectories by developing variations of Markov models~\cite{pirolli1999distributions,narvekar2015predicting,deshpande2004selective,mabroukeh2009semantic,manavoglu2003probabilistic}, n-gram sequence models~\cite{pitkow1999mininglongestrepeatin,su2000whatnext}, or hybrid models combining Markov models with ANNs~\cite{awad2007web} or clustering and association rule mining~ \cite{khalil2009integrated}. Our goal is not to develop a prediction algorithm, but to estimate the theoretical upper limit of predictability given users' sequence of web accesses. On comparing with prior work, we observe that the previously reported performance matches well with our estimated theoretical limits.

\section{Predictability Measurement Framework}\label{sec:framework}
In this section, we describe the framework for estimating the predictability of people's mobility in cyberspace from their web browsing traces. First, we outline how to construct trajectories from raw browsing data for each user. Then, we detail how we measure uncertainty in users' trajectories using concepts of entropy from the field of Information Theory. These concepts help us to uncover the extent to which individuals follow routines on the Web. Finally, we describe the procedure to estimate the theoretical predictability limits of a user's browsing behavior. 

\subsubsection{Sharing code} In this paper, we attempt to make the framework accessible to diverse research communities, and as such, our code is available as a Python library with corresponding online tutorial notebooks~\cite{webtrackinglibrary}.

\subsection{Constructing user trajectories}\label{sec:framework-trajectory}

\begin{figure}[b!]
    \centering
    \includegraphics[width=3.3in]{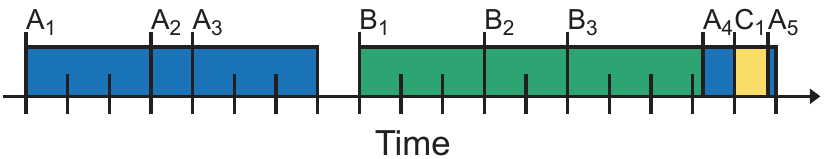}
    \caption{An illustration of raw web tracking data: A user visits nine different website URLs ($A_1$, $A_2$, $A_3$, $A_4$, $A_5$, $B_1$, $B_2$, $B_3$, and $C_1$, ) along the timeline. Each URL is from a domain (\ie $A$, $B$, and $C$) and belongs to a category (\eg search engine, social media site). Each visit has a time duration, and visits are non-overlapping. }
    \label{fig:trajectory_example}
\end{figure}

To study the mobility of users on the Web, we first need to construct their cyberspace trajectories from the raw web tracking data. Figure~\ref{fig:trajectory_example} illustrates a toy example of raw data: a user visits nine different website URLs, spending different amounts of time on each of them. In this example, each URL belongs to a domain (i.e., $A$, $B$, $C$), and they can be associated with a category, such as `news', `social media', or `search engine'. In our paper, we define a \emph{trajectory} as a discrete sequence of locations visited by a user, where a \emph{location} can be a website URL, a domain, or a category.

We propose two types of web trajectories: stationary and non-stationary. The former accounts for the amount of time spent on locations, while the latter ignores it. These trajectories enable us to study the role of time in measuring users' predictability.
In what follows, we describe each trajectory type using the toy data (Figure~\ref{fig:trajectory_example}) as the raw browsing trace data. Note that though most of the following examples are trajectories of domains, the definitions are generalizable to other types of locations (\ie categories or website URLs). 


\subsubsection{Series of time bins} 
The \textbf{stationary trajectories} ${T}^{\text{stat}}$ incorporate the amount of time a user spends on each location. When we analyze users' predictability using this trajectory, we are interested in the task of {\it predicting the location visited in the next time step}. Practical applications of such a task include real-time recommendations, dynamic caching, and dynamic advertising. With this type of trajectory, we can investigate the impact of visit duration on the emergence of users' routines on the Web. 

To generate this trajectory, we use a non-overlapping time window of size $\Delta t$ to create bins across time. Within each time bin, we choose the most visited location (\ie the location visited for the longest duration). We break ties randomly and ignore the time bins without any browsing activity. For example, using our toy data, the stationary trajectory of domains would be as follows:
\begin{equation*}
\begin{split}
{T}^{\text{stat}} = & \, \{A,A,A,A,A,A,A,B,B,B,B,B \\
& \, B,B,B,A,C\}.
\end{split}
\label{eq:binned_stationary}
\end{equation*}
We can observe that stationarity is captured in the trajectory. This method of generating trajectories is consistent with prior studies of human mobility~\cite{song2010limits,ikanovic2017alternative}.
 
\subsubsection{Series of visited locations}
The non-stationary trajectories ignore the amount of time a user spends on each location, ignoring users' tendency to stay longer at specific locations. When we analyze the predictability of non-stationary trajectories, we are interested in the task of {\it predicting the next visited location, irrespective of how much later}. Examples of applications include learning visitation patterns to automate users' web activities (\eg booking a holiday, or researching and buying products or services). These trajectories enable us to investigate users' routines that emerge regardless of the duration of visits. 

We build non-stationary trajectories using two different temporal schemes: binned and sequential. The former considers locations for each time bin, whereas the latter considers locations as they appear in the raw browsing trace data. These trajectories are defined as follows:

\begin{itemize}
    \item \textbf{Non-stationary binned trajectory} ${T}^{\text{binNonStat}}$:
    First, we create a stationary time series from the raw data using a time window of width $\Delta t$. Then, we compress this trajectory by removing repetitive adjacent locations from it. For instance, using our toy example, the non-stationary binned trajectory of domains looks as follows:
    \begin{equation*}
        {T}^{\text{binNonStat}} = \{A,B,A,C\}.
    \label{eq:binned_nonstationary}
    \end{equation*}
    
    By comparing this sequence with ${T}^{\text{stat}}$, we see that we have removed the stationarity from the trajectory. This method of trajectory construction is consistent with prior work~\cite{ikanovic2017alternative}.
    
    \item \textbf{Non-stationary sequential trajectory} ${T}^{\text{seqNonStat}}$:
    Instead of compressing the stationary trajectory, we compress the sequence of distinct locations from the raw data. 
    For example, applying this method to our toy data would generate the following trajectory of domains:
    \begin{equation*}
        {T}^{\text{seqNonStat}} = \{A,B,A,C,A\}. 
    \label{eq:domain_aggregated}
    \end{equation*}
    This approach creates trajectories that include all distinct locations that were visited independently of the amount of time spent on them.  For instance, in our example, domain $A$ appears at the end of trajectory ${T}^{\text{seqNonStat}}$, but not in ${T}^{\text{binNonStat}}$. 
    

\end{itemize}

Note that the two types of non-stationary trajectories converge to each other as $\Delta t$ gets smaller and smaller, becoming more and more independent of $\Delta t$.

\subsection{Estimating predictability of users' mobility on the Web}\label{sec:framework-predictability}

Having constructed a user's trajectory, we can now study the dynamics of the user's mobility on the Web. Without loss of generality, we can represent any trajectory as a discrete series  $$T = \left\{x_1, x_2, \hdots, x_\ell\right\},$$ where $x_t \in \mathcal{V}$ is a location (i.e., a website URL, a domain, or a category), $\mathcal{V}$ is the set of visited locations, and $\ell$ is the trajectory length. Here we are interested in the dynamics of this time series; we want to study (1) the location preferences of a user and (2) the emerging visitation patterns in these preferences. Specifically, we aim to investigate the uncertainty embedded in this time series. 

In this paper, we use three concepts of uncertainty from the Information Theory field: 
\begin{enumerate}[(i)]
    \item Time-uncorrelated entropy ($S^\text{unc}$) that accounts for the frequency with which users visit locations. \item Maximum entropy ($S^\text{rand}$) that considers only the number of unique locations visited. 
    \item Time-correlated entropy ($S$) which considers the visitation patterns of the users. 
\end{enumerate}
These three measures allow us to quantify the uncertainty in users' browsing behavior, enabling us to estimate the extent to which users exhibit routineness and compute the theoretical limits of predictability of users' trajectories.

\subsubsection{Users' location preferences}
To study a user's location preferences, we can examine the number of unique locations visited by the user and the probability $p(i)$ of visiting a location $i$ (i.e., the normalized frequency of visits). 

The size of $\mathcal{V}$, denoted by $N=|\mathcal{V}|$, relates to the preference breadth of a user. It tells us how broadly this user explores the Web throughout a particular time frame. This quantity, however, neglects the frequency with which the users visit locations, missing the likely existence of favorite websites, categories, or domains. To account for frequency, we need to examine the spread of the probability distribution $p(i)$. For instance, if a user visits all locations at the same frequency, then $p(i) \approx 1/N$, whereas if a user mostly visits a specific location $j$, then $p(i)$ peaks at $i=j$. We want to quantify the peakiness of the distribution $p(i)$. 

From an information-theoretic perspective, we want to measure the uncertainty of a random variable (in our case, the random variable is the time series $T$). For analyzing this scenario without accounting for temporal correlations, we can use Shannon entropy, denoted here as the \textbf{time-uncorrelated entropy} $S^\text{unc}$, defined as
\begin{equation}
S^\text{unc}(p) = -\sum_{i\in \mathcal{V}} p(i) \log_2 p(i)
\label{eq:Shannon}
\end{equation}
and expressed in bits~\cite{Cover2006}. $S^\text{unc}(p)$ measures the average amount of memory needed to store the outcome of the random variable associated with $p(\cdot)$. 
That is, the more certain we are about the outcome of a random variable, the less memory we need to store the variable, thus the lower $S^\text{unc}(p)$.

In our case, Eq.~(\ref{eq:Shannon}) quantifies the uncertainty concerning the locations a specific user visits over time. For instance, when a user keeps visiting only one location, the entropy $S^\text{unc}(p)$ is zero because of the low uncertainty of this user's behavior. In contrast, a user without a favorite location exhibits the highest uncertainty and the maximum entropy. 
The \textbf{maximum entropy} $S^{\text{rand}}$ occurs when $p(i) = 1/N$ for $\forall i$ and can be defined as the following: 
\begin{equation}
    S^{\text{rand}}(p) = \log_2 N.
\end{equation}
This quantity represents the uncertainty in a user's location when the user's visitation behavior follows a uniform distribution (\ie without any preference towards specific locations). $S^{\text{rand}}(p)$ can be seen as the worst-case scenario of a variable with a sample space of size $N$, such that this uncertainty equals to guessing a user's location \textit{randomly}. 

 \begin{figure}[b!]
    \centering
    \includegraphics[width=3.3in]{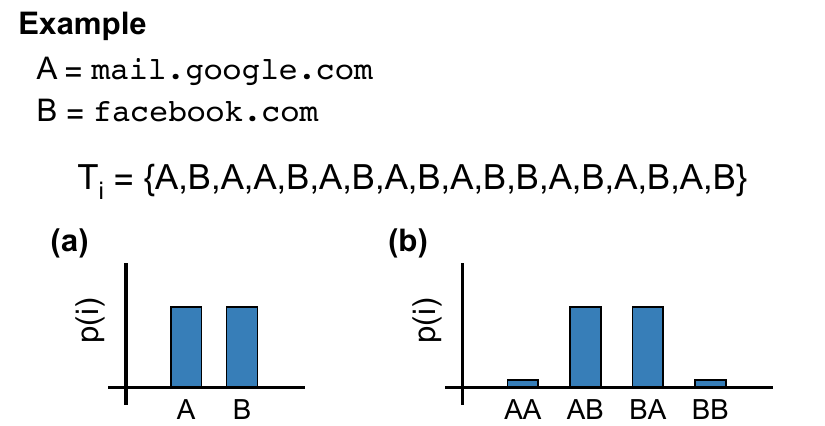}
    \caption{
We need to account for long-range temporal correlations to study the predictability of users' browsing behavior. For example, consider a hypothetical user trajectory of accessing an e-mail account then a social media website (top). When a user follows such a sequence over time, the absence of favorite locations makes the Shannon entropy to approach its maximum value, $S^\text{unc}\approx S^{\text{rand}}$ \textbf{(a)}. This trajectory, however, is quite predictable; we can see a clear pattern repeating itself. Indeed, when we examine the probability of the blocks of length $L=2$, we find some favorite patterns in the trajectory \textbf{(b)}. The rate at which the uncertainty increases with $L$ is called the source entropy rate. This quantity measures the intrinsic uncertainty of a user trajectory as we learn about the sequences that this user follows over time.}
\label{fig:entropy_toy}
\end{figure}

\subsubsection{Users' visitation patterns}
Though $S^\text{unc}(p)$ provides us with a simple uncertainty measure, it neglects long-range temporal correlations in the trajectories---a potential feature in web mobility.  A temporal correlation might emerge because of repetitive patterns of visitation on the Web (e.g.,~always accessing a social media website after reading the news). Such patterns in a user's trajectory decrease the uncertainty about this user, regardless of the Shannon entropy. Indeed, a user might lack favorite locations but have favorite patterns of visitation (see Figure~\ref{fig:entropy_toy}). To capture this uncertainty in web mobility, we need to examine trajectories as the result of processes that generate sequences. 

First, we denote a block of $L$ consecutive random variables as $X^L=X_1\hdots X_L$, and $x^L=x_1\hdots x_L$ is an instance of this variable. In our case, $x^L$ can be any sequence of length $L$ in a trajectory of a user. If we denote $p^L(\cdot)$ as the probability function of $X^L$, we can define the entropy of $p^L$ as the following: 
\begin{equation}
S'(p^L) = -\sum_{s^L\in \mathcal{V}^L} p^L(s^L) \log_2 p^L(s^L),
\label{eq:total_Shannon}
\end{equation}
where $\mathcal{V}^L$ is the sample space containing all the visited sequences of length $L$, and we establish $S'(p^0) = 0$~\cite{Crutchfield2003}. This quantity is the so-called block entropy or total Shannon entropy of length-$L$, and it is a convenient measure of uncertainty of the sequences in a trajectory~\cite{Crutchfield2011}. 

Instead of focusing on a specific length $L$, we can examine the relationship between $S'(p^L)$ and $L$ to investigate the uncertainty in the trajectories. Specifically, we would like to analyze how $S'(p^L)$ changes as we have longer and longer $L$-blocks. In this way, we learn about the uncertainty of a trajectory while discounting for the recurrent patterns in it. For this purpose, we use the source entropy rate of a stochastic process, defined as 
\begin{equation}
h_\mu = \lim_{L\to \infty} {S'(p^L)}/{L},
\end{equation}
when the limit exists~\cite{Crutchfield2003,Cover2006}. The entropy rate tells us the irreducible uncertainty of a process that persists even when we account for the temporal correlations. 
In our case, $h_\mu$ quantifies the intrinsic uncertainty of a trajectory even when we learn about the sequences (i.e., patterns of visitation) that this person follows throughout time. 

In our paper, we follow previous works~\cite{song2010limits,Smith2014,ikanovic2017alternative} and estimate the
entropy rate of a trajectory using the Lempel--Ziv compression algorithm; we call this estimate the \textbf{time-correlated entropy} and denote it as {\bf $S$}.
The Lempel--Ziv algorithm attempts to find the optimal dictionary to compress a given sequence, and it has been shown to quickly converge to the entropy rate as the sequence length approaches infinity~\cite{Kontoyiannis1998}.

By comparing time-correlated entropy with time-uncorre\-lated entropy, we can investigate the role of routines in users' web browsing behavior. The higher the difference between the entropies, the more individuals repeat visitation patterns. These recurrent visitation patterns occur because of preferred sequences in visiting websites, categories, or domains. We call this preference \textbf{web routineness}. This property of web browsing behavior emerges from individuals' choices under the structural constraints of the Web and can affect 
the extent to which we can predict their trajectories.

\subsubsection{Estimating predictability}
These entropy quantities enable us to assess the uncertainty embedded in the browsing behavior of individuals; however, they fail to explicitly tell us how well we could predict a user's trajectory despite its intrinsic uncertainty. We want to learn about the probability $\Pi$ of correctly predicting future locations, given a past series of observations. It is possible to show that $\Pi$ is subject to Fano's inequality\footnote{We omit details of the analytical derivation and refer the interested reader to the literature~\cite{song2010limits,Smith2014}.}, and has an upper bound, denoted as $\Pi^{\text{max}}$.
 
In our case, this upper bound reveals the theoretical upper limit to predict an individual's future location correctly if we restrict ourselves to only the browsing trajectories from the web tracking data. For instance, a user having an upper limit of  $\Pi^{\text{max}}=0.6$ exhibits an intrinsic uncertainty that makes their behavior indistinguishable from random $40\%$ of the time. Because of this randomness, the best accuracy level that a predictive algorithm can achieve for this user is $60\%$. 
That is, $\Pi^{\text{max}}$ tells us how much predictive power we can theoretically have by possessing just the browsing trajectories from the web tracking data of an individual. 


It is impractical to compute $\Pi^{\text{max}}$ directly, but the quantity has an explicit relationship with the entropy rate:
\begin{equation}
S = H_b(\Pi^{\text{max}}) + (1 - \Pi^{\text{max}})\log_{2}(N-1),
\label{eq:predict}
\end{equation}
where $H_b(\cdot)$ is the binary entropy function, defined as $H_b(p) = -p \log_{2} p + (1-p) \log_{2} {(1 - p)}$, and $N$ is the alphabet size (i.e., the number of unique locations visited)~\cite{song2010limits}. To find the \textbf{predictability upper-bound} $\Pi^{\text{max}}$ of a user, we just have to solve Eq.~(\ref{eq:predict}) using a numerical solver, provided that we know $S$ and $N$ for this user. 
This approach also allows us to find the hypothetical predictabilities of users if their trajectories were random or absent of temporal correlations by substituting $S$ in Eq.~(\ref{eq:predict}) with the corresponding entropy value. More precisely, we replace $S$ with $S^{\text{rand}}$ and $S^{\text{unc}}$ to find { $\Pi^{\text{rand}}$} and { $\Pi^{\text{unc}}$}, respectively.

\subsubsection{Comparing predictabilities}
Having the three predictability values ($\Pi^{\text{rand}}$, $\Pi^{\text{unc}}$, and $\Pi^{\text{max}}$) for the users provides us with a framework for comparing users and investigating the impact of emerging preferences and routines in their trajectories. First, predictability values lie within $[0, 1]$, making it convenient to compare users with a varying number of visited \mbox{locations---different} from absolute entropy numbers. Second, the distinct types of predictability help us interpret the results by yielding two null models.
Indeed, finding the hypothetical values of $\Pi^{\text{rand}}$ and $\Pi^{\text{unc}}$ helps us to contextualize $\Pi^{\text{max}}$. Each of these predictability values represents a version of a user's trajectory when we remove a specific feature of it. $\Pi^{\text{unc}}$ represents the predictability of a trajectory if we randomize the original trajectory, removing all the temporal correlations. $\Pi^{\text{rand}}$ represents the predictability of a trajectory if we remove all repeated visits to locations from the original trajectory, removing all user preferences. With these values as a baseline, we can now compare them with $\Pi^{\text{max}}$ 
to understand the impact of web routineness on the predictability of a user's web browsing behavior.

\section{Dataset of Web Browsing Traces}



\begin{table*}[t]
\centering
 \caption{Comparison of German population margins of gender (male, female) and age (18-80 years) with our sample's composition (percentage values). Filtered sample in parentheses. Unfiltered sample refers to all $2,148$ users, while the filtered sample depicts the $1,455$ users who have long enough trajectories to estimate predictability reliably (see Section~\ref{sec:sufficient_data}).}
\label{tab:genderagedist}
\begin{tabular}{@{}lcclccclccc@{}}
\toprule
\textbf{Gender} & \textbf{Population} & \textbf{Sample} &  & \textbf{Age} & \textbf{Population} & \textbf{Sample} &  & \textbf{Age} & \textbf{Population} & \textbf{Sample} \\ \cmidrule(r){1-3} \cmidrule(lr){5-7} \cmidrule(l){9-11} 
Female          & 50.6                & 51.2 (48.5)     &  & 18-24        & 9.7                & 12.5 (11.1)     &  & 45-54        & 19.3                & 22.4 (24.5)     \\
Male            & 49.4                & 48.8 (51.5)     &  & 25-34        & 16.3                & 19.8 (18.4)     &  & 55-64        & 18.6                & 21.2 (23.8)     \\
                &                     &                 &  & 35-44        & 15.6                & 18.1 (15.9)     &  & 65-80        & 20.5                & 4.4 (4.8)       \\ \bottomrule
\end{tabular}
\end{table*}

Our anonymized dataset consists of one month's (October 2018) web tracking data of $2,148$ German users. 
For each user, the data contains the URL\footnote{URLs are anonymized, and personally identifiable information such as user names or passwords have been removed.} of the webpage the user visited, the domain of the webpage, category of the domain\footnote{Categories were inferred using \url{https://docs.webshrinker.com/v3/website-category-api.html}, which provides 41 distinct categories. We manually added two more categories `email' and `productivity'. Moreover, for popular domains with many sub-domains with distinctly different functions (such as \textit{google.com}), we manually re-coded the sub-domains to correct categories. 
See \url{https://github.com/gesiscss/webtracking/} for more details.}, time of visit, and active seconds spent by the user on the page. In total, these 2,148 users made 9,151,243 URL visits, spanning 49,918 unique domains.

\subsubsection{Evaluating web tracking data} We used web traffic data from Alexa\footnote{While Alexa is also not representative of German online population, we use it since it is the most established source for web traffic data: \url{https://docs.aws.amazon.com/AlexaTopSites/latest/ApiReference_TopSitesAction.html}.} to evaluate whether the most visited domains in Germany also feature prominently in our dataset.
We compared the 5000 most visited domains from Alexa with the aggregated number of visits to these domains in our web tracking data. We found them to be highly correlated (Pearson's correlation coefficient $r$ = 0.78). 



\subsubsection{Comparing the demographics of our panelists and the German population}
For each user in our dataset, we have self-reported information (collected via a survey) about their gender and age. We compare our sample's distributions with German population margins for gender
and age (Table~\ref{tab:genderagedist}).\footnote{In the absence of high-quality data about German internet users' demographics, we compare with offline German population margins for the year 2018: \url{https://www.destatis.de/EN/Themes/Society-Environment/Population/Current-Population/_node.html}.}
We observe that our sample's gender distribution matches closely with the German population's. However, the older population (65 years and above) is under-sampled in our panel. This difference could potentially be due to lower internet usage by older people, or due to the opt-in, non-probability recruitment of the panelists.
\subsubsection{Ethics}
The web tracking data for the users in our dataset was collected by a GDPR compliant digital panel company in Europe. All the participants of the company's online panel have agreed to anonymously share their survey (\eg the gender and age information we use) and behavioral data for business or scientific purposes. 
The participants willing to be tracked install an executable file on their desktop. In return, they receive a monthly reward for their data.

\subsubsection{Sharing data} We share our anonymized dataset
to spur further research in this area~\cite{webtrackingdata}.





\section{Guidelines for Applying the Predictability Measurement Framework}\label{sec:design}

We now provide some practical guidelines for applying the framework. As the first step, we present a principled method for ensuring that the data is sufficient for measuring predictability. Then, we discuss the impact of some design decisions on the measured predictability and offer guidelines that could help researchers to reflect on these decisions systematically.

\subsection{Ensuring data is sufficient to estimate the  predictability}\label{sec:sufficient_data}

\begin{figure}[b!]
    \centering
    \includegraphics[width=1.02\linewidth]{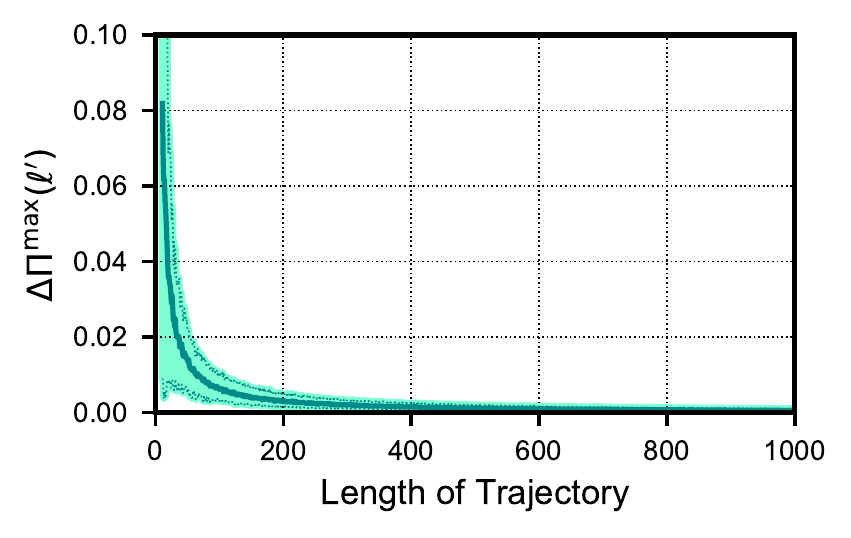}
    \caption{ Mean absolute change in maximum predictability with increasing trajectory length for random 100 users. The shaded region depicts the $5^{th}$ to $95^{th}$ quantile range.} 
    \label{fig:change_max_predictability_with_increasing_trajectory}
\end{figure} 

To ensure that we have enough data to estimate users' predictability, we examine the convergence of $\Pi^\text{max}$ as we add more and more data. The rationale here is that the quality of $S$ (i.e., the estimate of the entropy rate, needed to find $\Pi^\text{max}$) depends on the trajectory length. As this length approaches infinity, $S$ converges to the entropy rate. We study this convergence by analyzing the predictability of trimmed versions of users' trajectories. For this, we first denote $\Pi^\text{max}(\ell')$ as the predictability of a user estimated using only the first $\ell'$ locations of this user's trajectory; then, we study the absolute consecutive differences of $\Pi^\text{max}(\ell')$, defined as:
\begin{equation*}
    \Delta \Pi^\text{max}(\ell') = |\Pi^\text{max}(\ell') - \Pi^\text{max}(\ell'-1)|,
\end{equation*}
where $1<\ell'\leq \ell$.

To examine the predictability convergence, we selected $100$ random users and computed $\Delta \Pi^\text{max}$ of their non-stationary $T^\text{binNonStat}$ domain trajectories (since these are the shortest trajectories). We found that the predictability values stabilize for a trajectory length of $\ell' = 100$, such that the mean difference is small ($\Delta \Pi^\text{max}(100) = 0.006$), as shown in Figure~\ref{fig:change_max_predictability_with_increasing_trajectory}. Therefore, we decided that the length of users' non-stationary trajectories must be at least $100$ locations long. In the rest of the paper, we perform our analysis with the $1,455$ users who satisfy this criterion, excluding $693$ users from our study. These $1,455$ users made a total of $8,910,779$ URL visits spanning $48,819$ unique domains, and their gender and age distributions are depicted in Table~\ref{tab:genderagedist}.


\subsection{Impact of design decisions on predictability}\label{sec:impact_design_dec}


Next, we discuss the impact of three critical design decisions on the estimated predictability: (i)~stationarity of trajectories, (ii) temporal resolution, and (iii) spatial resolution.

\subsubsection{Impact of stationarity in trajectories}
\begin{figure}[h]
    \vspace{-1mm}
    \centering
    \includegraphics[width=\linewidth]{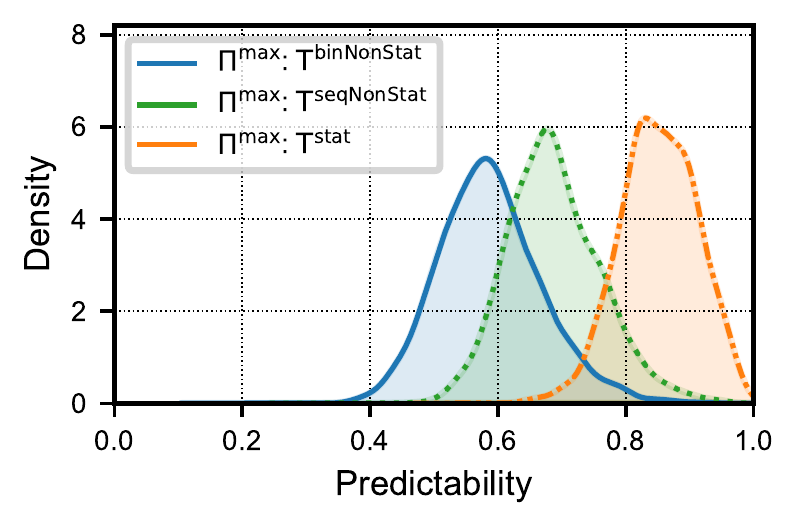}
    \caption{Impact of stationarity in the trajectories on the predictability.}
    \label{fig:trajectory_stationarity}
    \vspace{-3mm}
\end{figure}

When applying the predictability framework, we first need to decide which trajectory to construct. This decision is primarily driven by which prediction task we want to study.
While ${T}^\text{stat}$ is suitable for predicting the next visited location in the next time bin,  ${T}^\text{binNonStat}$ or ${T}^\text{seqNonStat}$ are suitable for predicting the next visited location independent of when it is visited. To investigate how predictability changes depending on these tasks, we computed the predictability of the users using these three types of trajectory. We found that these tasks satisfy the following inequality:
$${\Pi}^\text{max}_\text{binNonStat} < {\Pi}^\text{max}_\text{seqNonStat} < {\Pi}^\text{max}_\text{stat},$$
where the predictability values correspond to the trajectories ${T}^\text{binNonStat}$, ${T}^\text{seqNonStat}$, and ${T}^\text{stat}$, respectively. We observe that as the stationarity in the trajectories increases (i.e., the user stays in the same location for more consecutive time bins), the maximum predictability also increases, as shown in Figure~\ref{fig:trajectory_stationarity}. Our results indicate that it is easier to predict the location visited by the user in the next time bin rather than the next (distinct) location visited irrespective of the time of visit.

\subsubsection{Impact of temporal resolution}
The next decision is the choice of temporal resolution $\Delta t$. 
We expect that when $\Delta t$ is close to zero, the uncertainty in the next visited location (\ie entropy) is minimal, and the predictability approaches $100\%$. To investigate this conjecture, we computed the entropy and predictability of the users' $T^\text{stat}$ domain trajectories using varying time bins $\Delta t \in$ [$0.25$, $0.5$, $0.75$, $1$, $2$, $3$, $4$, $5$, $6$, $7$, $8$, $9$, $10$, $11$, $12$, $13$, $14$, $15$] minutes.\footnote{We found the average session length to be $883$ seconds, which is just under $15$ minutes. Hence, we vary $\Delta t$ up to $15$ minutes only.} 

\begin{figure}[h!]
    \centering
    \includegraphics[width=\linewidth]{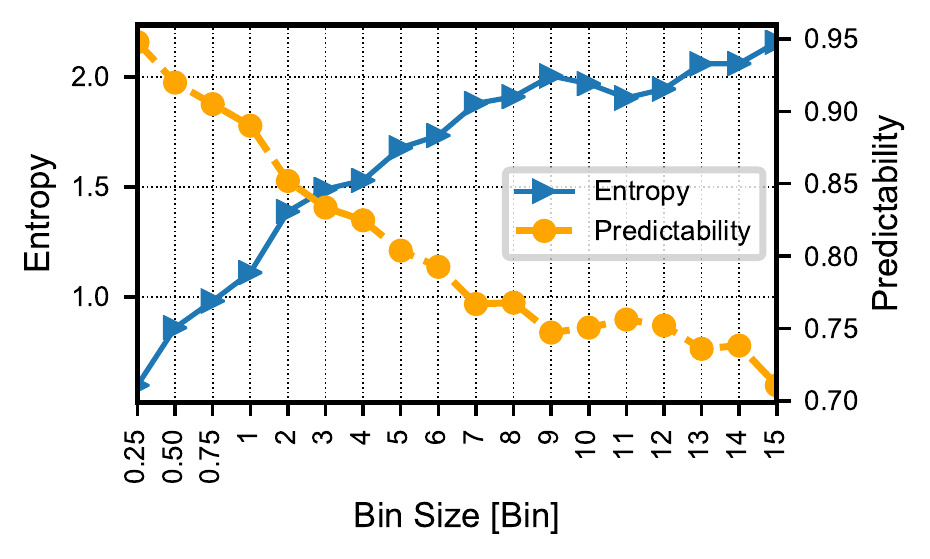}
    \caption{Impact of temporal resolution on the predictability.}
    \label{fig:temporal_resolution}
    \vspace{-3mm}
\end{figure}

As expected, we found that, as the temporal resolution gets more coarse-grained, the entropy increases while the predictability decreases (see Figure~\ref{fig:temporal_resolution}).
Note that different values of $\Delta t$ represent slightly different prediction tasks.

In our case, we would like to set a reasonable value to $\Delta t$ based on our data. For this, we measure the average amount of time that users spend on each domain visit, finding an average of around $30$ seconds (M=36.91, 95\% CI [35.53, 38.29]). Note that the larger the size of $\Delta t$, the more short-duration visits would get removed from trajectories. Therefore, we use $\Delta t = 1$ minute for all other results in the paper.\footnote{We also generated our results with $\Delta t$ equal to $3$, $5$, and $8$ minutes, finding that the trends in our results still hold.} 

\begin{figure*}[t]
\begin{tabular}{ccc}
  \includegraphics[width=0.3\linewidth]{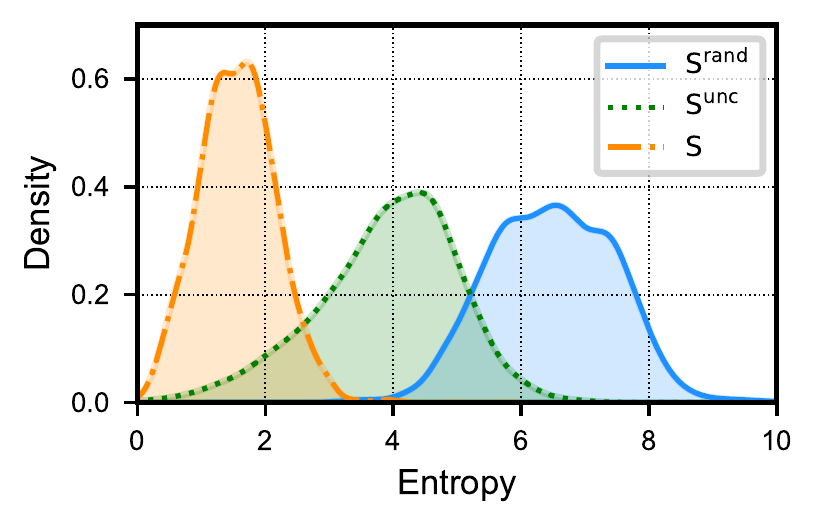} & \includegraphics[width=0.3\linewidth]{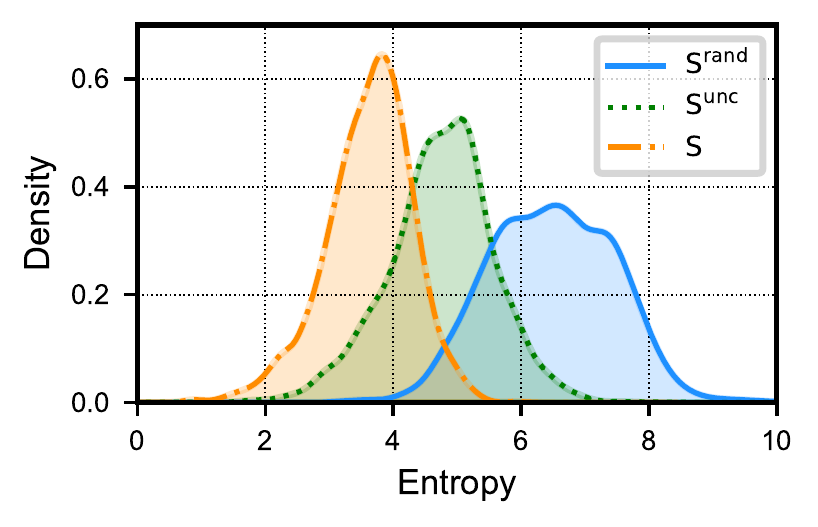} &  \includegraphics[width=0.3\linewidth]{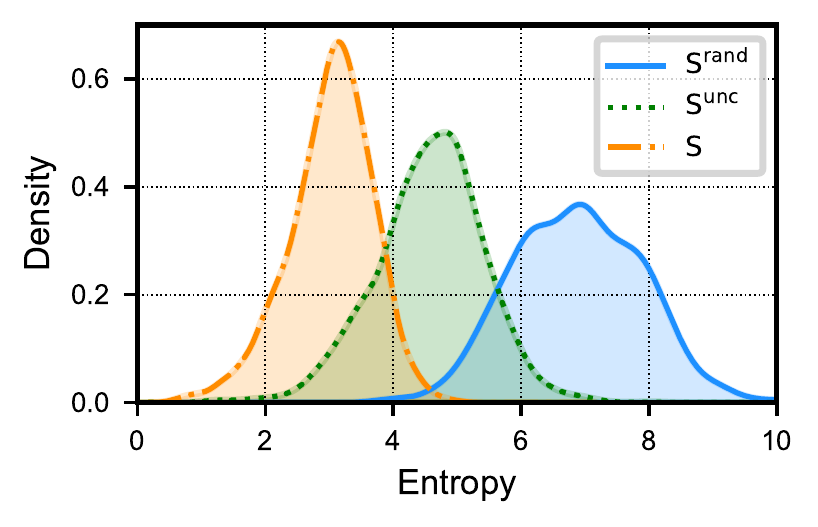} \\ \includegraphics[width=0.3\linewidth]{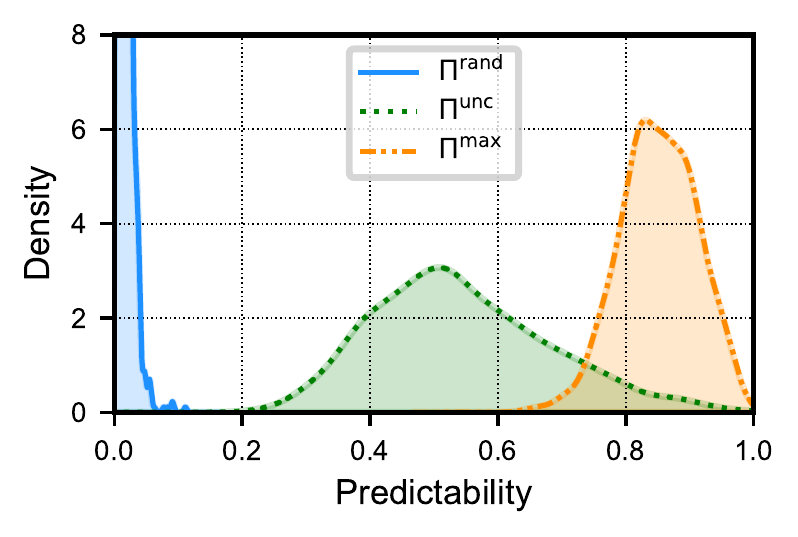} &  \includegraphics[width=0.3\linewidth]{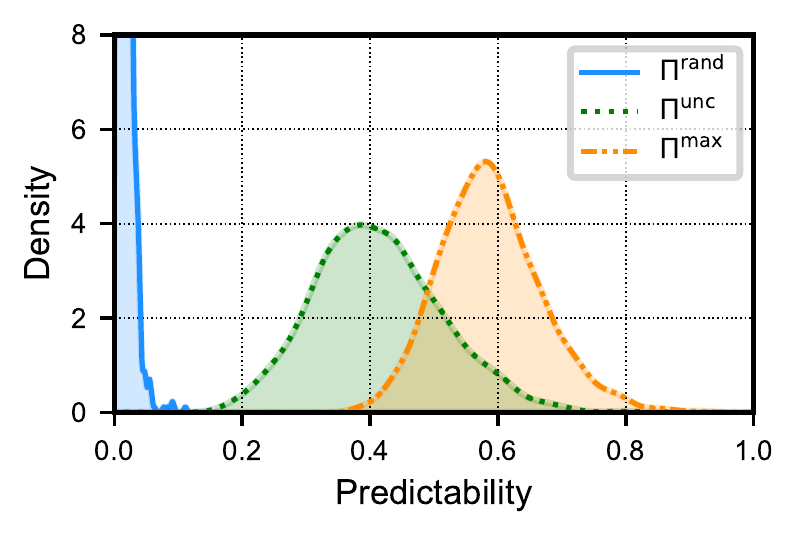} &  \includegraphics[width=0.3\linewidth]{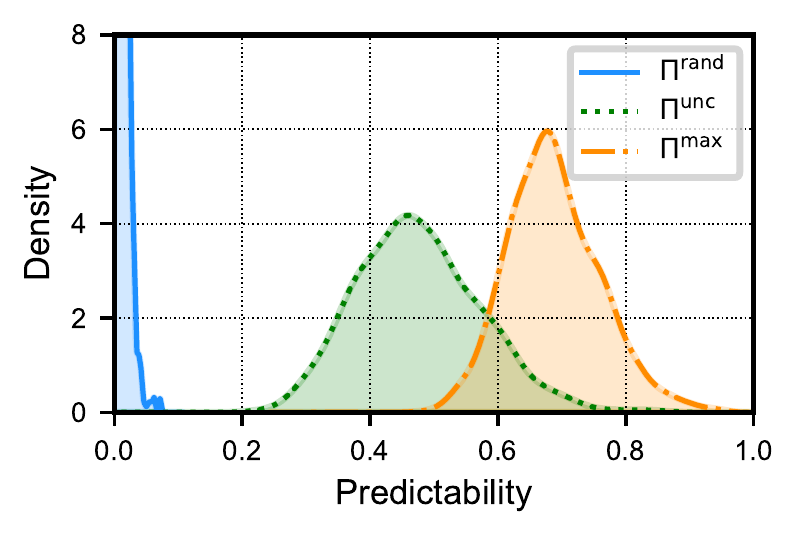} \\
   & & \\
  (a) Stationary  & 
  (b) Non-stationary binned  &
  (c) Non-stationary sequential \\
   \quad\quad trajectory $T^{\text{stat}}$  & trajectory  $T^{\text{binNonStat}}$&  trajectory $T^{\text{seqNonStat}}$ \\
\end{tabular}
\caption{Distribution of the three types of entropies ($S^{\text{rand}}$ in blue, $S^{\text{unc}}$ in green, and $S^{\text{}}$ in orange in the upper panel), and predictabilities ($\Pi^{\text{rand}}$ in blue, $\Pi^{\text{unc}}$ in green, and $\Pi^{\text{max}}$ in orange in the lower panel) of all users in our dataset computed using:\textbf{ (a)} stationary trajectories $T^{\text{stat}}$, \textbf{(b)} non-stationary binned trajectories $T^{\text{binNonStat}}$, and \textbf{(c)} non-stationary sequential trajectories $T^{\text{seqNonStat}}$.} 
\label{fig:entropy-predictability-overall}
\end{figure*}

\subsubsection{Impact of spatial resolution}
\begin{figure}[b!]
    \centering
    \vspace{-3mm}
    \includegraphics[width=\linewidth]{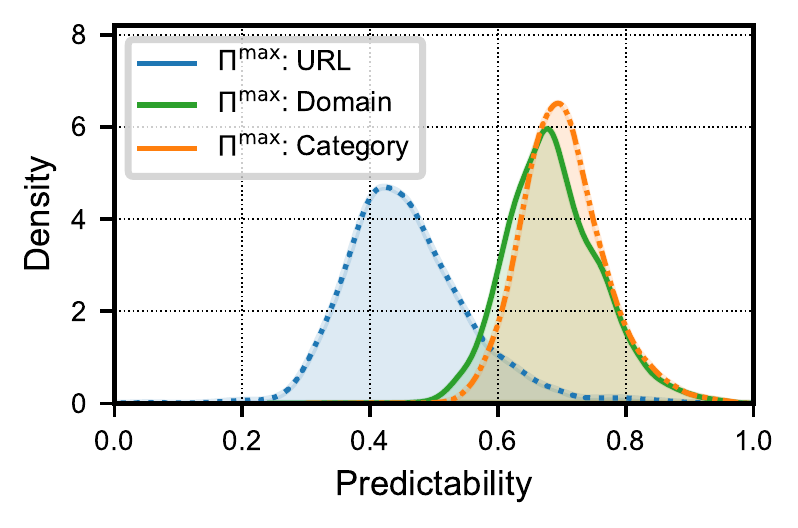}
    \caption{Impact of spatial resolution on the predictability.}
    \label{fig:spatial_resolution}
\end{figure}

The final design decision pertains to the spatial resolution of the locations in the analysis. In the cyberspace, this spatial resolution relates to hierarchical features of locations, such as URLs, sub-domains, domains, and categories. Deciding the resolution of the analysis typically depends on the problem at hand. In the limiting case, when the spatial resolution is too coarse-grained, and every web location is virtually the same, the predictability of the next visited location is perfect (i.e., the next location is always the same location).
To investigate the role of spatial resolution, we measure the predictability of users' $T^\text{seqNonStat}$ trajectories of locations with decreasing spatial resolution, from URLs ($2,910,824$ unique URLs) to domains ($48,819$ unique domains) to categories ($409$ unique categories). We find that, indeed, the maximum predictability increases with more coarse-grained spatial granularity, as shown in Figure~\ref{fig:spatial_resolution}. These results imply that we should only compare the predictability of users' movements if the spatial resolution is similar. In the rest of the paper, we present results at the resolution of domains, unless specified otherwise.

 \begin{figure*}
\begin{tabular}{cc}
  \includegraphics[width=0.47\linewidth]{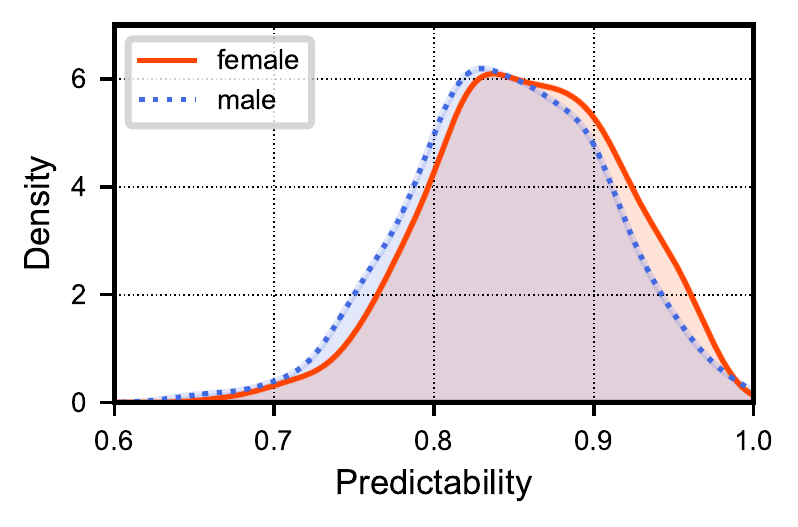} &   \includegraphics[width=0.47\linewidth]{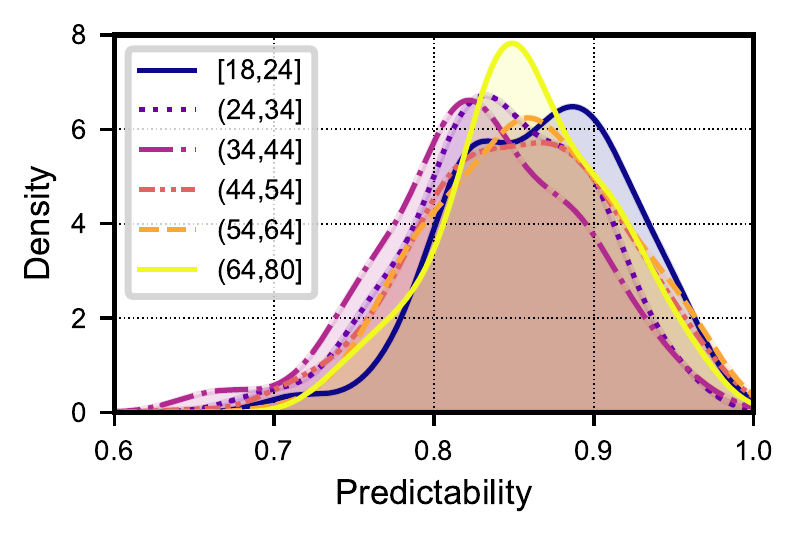} \\
   & \\
  (a) Gender  & (b) Age \\
\end{tabular}
\caption{Distribution of predictability $\Pi^{\text{max}}$, computed using stationary trajectories $T^{stat}$, for users in our dataset grouped by: \textbf{(a)} gender and \textbf{(b)} age.}
\label{fig:predictability-gender-age}
\end{figure*}

\section{Predictability of Users'  Mobility on the Web}

For each user in our dataset, we computed the three entropy ($S^{\text{rand}}$, $S^{\text{unc}}$, and $S^{\text{}}$) and predictability ($\Pi^{\text{rand}}$, $\Pi^{\text{unc}}$, and $\Pi^{\text{max}}$) values defined in Section~\ref{sec:framework-predictability} for the stationary ($T^\text{stat}$) and non-stationary ($T^\text{binNonStat}$, $T^\text{seqNonStat}$) trajectories of visited domains; their distribution are presented in Figure~\ref{fig:entropy-predictability-overall}. 

For both types of trajectories, we found that the `real' uncertainty ($S$) in people's movement in cyberspace is much lower than the expected entropy values when we disregard the order ($S^\text{unc}$) or users' preferences ($S^\text{rand}$) for the visited locations, as shown in the upper panel of Figure~\ref{fig:entropy-predictability-overall}.
These results indicate that people follow repetitive visitation patterns on the Web, that is, people exhibit web routineness in their browsing trajectories.

In the case of the theoretical limits of predictability, we found that the more information we have about the users' trajectories---from the number of locations ($\Pi^{\text{rand}}$) to preference for locations ($\Pi^{\text{unc}}$) to order of location visited ($\Pi^{\text{max}}$)---the higher the predictability. 
Our results show that the mean predictability increases from less than $5$\% to more than $85$\%
($95$\% CI $[84.79, 85.41]$) 
for $T^\text{stat}$, around $60$\% 
($95$\% CI $[58.38, 59.20]$)
for $T^\text{binNonStat}$, and around 70\%
($95$\% CI $[68.40, 69.14]$) 
for $T^\text{seqNonStat}$ (see the lower panel of Figure~\ref{fig:entropy-predictability-overall}).
Such an increase in predictability indicates that the routines users follow increase the predictive power we can theoretically achieve based on their web browsing trajectories.


When we compare our estimated theoretical limit of predictability with the performance of previously developed web access prediction algorithms that only utilize users' web access sequences,~\cite{awad2007web,narvekar2015predicting,deshpande2004selective,mabroukeh2009semantic,khalil2009integrated,manavoglu2003probabilistic,pitkow1999mininglongestrepeatin}, we observe that prior results are in line with our estimated theoretical limits.\footnote{We compare the prediction accuracy from prior work with our estimated limits for $T^\text{seqNonStat}$ trajectories since the web access sequences used in prior work match closest to them. However, we must note that without knowing the differences between the trajectories used in previous works and our trajectories as well as the differences in their server log datasets and our web tracking dataset, this is a less than optimal comparison.}

Lastly, note that even though there are definite peaks of the maximum predictability distributions in Figure~\ref{fig:entropy-predictability-overall}, there is a considerable spread across different users. This spread suggests that it is easier to predict the next visited location for some users than for the others.  
In the next section, we will explore which behavioral and demographic factors explain this variation in users' predictability.

\subsection{What explains the predictability?}

\subsubsection{Demographics}
We now turn our attention towards examining whether users with different demographic attributes demonstrate different degrees of maximum predictability for their web browsing behavior. We focus our analysis on two demographic features: gender and age. Although in this section, we present the results computed using $T^\text{stat}$ trajectories, our results also hold for non stationary trajectories.



\begin{itemize}
    \item {\bf Gender:}
     Our dataset of $1,455$ users consists of $51.5\%$ men and $48.5\%$ women (see Table~\ref{tab:genderagedist}). We evaluated the differences between the distributions of the predictability for the men and women (depicted in Figure~\ref{fig:predictability-gender-age}(a)) using a two-sample Kolmogorov-Smirnov (KS) test~\cite{berger2014kolmogorov}. We obtained a KS score of $0.086$ at the p-value of $0.008$ for the predictability. Since the p-value is below $0.05$, we can reject the null hypothesis and conclude that the distributions of predictability of men and women are significantly different. Additionally, we also computed the effect size using Cliff's delta~\cite{cliff1993dominance}. The value of $d=0.11$ indicates that there is an $11\%$ chance that a randomly chosen woman has higher predictability than a randomly chosen man. 

\item{\bf Age:}
For studying whether users of different ages demonstrate different browsing behavior and hence are predictable to differing extents, we divided our set of $1,455$ users into six age groups: [18-24, 25-34, 35-44, 45-54, 55-64, 64-80] which contain [161, 267, 231, 357, 347, 70] users respectively (see Table~\ref{tab:genderagedist}). Following the same procedure as above, we applied the two-sample KS test to the probability distribution of users in every pair of age groups.
As before, we also computed Cliff's delta. We show the pairs of age groups for which the differences were significant in Table~\ref{tab:predictability_age}.\footnote{To adjust for multiple pairwise comparisons ($n=15$), we apply the Bonferroni correction~\cite{shaffer1995multiple}. Doing so, we set alpha value for the entire set of age-based comparisons to $\alpha=0.05$ by setting alpha value of each comparison to $\alpha/n = 3.3 \times 10^{-3}$.}
We note that the ``middle-aged'' users (35-44) are typically lesser predictable than users of most other ages, as shown in the table.  Interestingly, even among the ``young'' users (18-24 and 25-34), $d=-0.24$ indicates a 24\% chance that a randomly chosen user below the age of 24 has higher predictability than a randomly chosen user above 24 years of age. We show the distribution of the predictability of users in different age groups in Figure~\ref{fig:predictability-gender-age}(b).
\end{itemize}


\begin{table}
\caption{Differences in the predictability of users of different age groups. (Only significantly different pairs included.)}
\label{tab:predictability_age}
\centering
\begin{tabular}{cccr}
\toprule
{\bf Age} & \multicolumn{2}{c}{\bf KS test} & \multicolumn{1}{c}{{\bf Cliff's}} \\
\cmidrule{2-3}
 {\bf groups} & { KS score}  & { p-value} & \multicolumn{1}{c}{{\bf delta}}\\
 \midrule
 18-24 \& 25-34 & $0.19$ & $8.3\times10^{-4}$ & $-0.24$ \\ 
 18-24 \& 35-44 & $0.26$ & $1.8\times10^{-6}$ & $-0.34$ \\
 35-44 \& 45-54 & $0.16$ & $6.2\times10^{-4}$ & $0.19$\\
 35-44 \& 55-64 & $0.21$ & $3.1\times10^{-6}$ & $0.25$\\
 35-44 \& 65-80 & $0.28$ & $2.1\times10^{-4}$ & $0.27$ \\
 \bottomrule
\end{tabular}
\end{table}


\subsubsection{Browsing Behavior}
Likely, the differences in predictability between men and women and young and older people can be explained by differences in their browsing behavior. We, therefore, analyze the impact of users' browsing characteristics on the predictability of the domain visited in the next time bin (using $T^{\text{stat}}$ trajectories) and analyze differences between the behavior of different groups. We observe similar trends for the next visited categories and other trajectories but omit the results for brevity. For this analysis, we constructed some augmented features for each user in our dataset based on their web browsing behavior. We then examined how the users' maximum predictability varies with these browsing features using Pearson's correlation coefficient $r$. 
Next, we describe our constructed browsing features and the variation of predictability with them.

\begin{itemize}
    \item {\bf User activity:}
    We quantify how active a user is via the total active seconds that the user is browsing and the count of the user's total domain visits during our one-month observation period. We find that predictability is positively correlated with total active seconds that the user spends browsing ($r=0.4001$); however, it is not correlated with the total domain visits ($r=0.0074$).
    
    \item {\bf Diversity of user interests:}
    We capture how varied user's interests are, as the number of distinct domains and categories that the user visits. We observe that the more varied a user's interests are, the lower is the user's predictability as indicated by a negative correlation between predictability and number of distinct domains ($r=-0.2282$) and categories ($r=-0.1528$) visited.
    
    \item {\bf User stationarity:}
    When a user spends more amount of time on average on each visit to a domain, we consider the user to have more stationarity in their browsing behavior since they are stationary on a domain for a long time before moving to another. Therefore, we consider the degree of stationarity in a user's browsing trajectory to be captured by the mean and median amount of time spent on each domain visit. In both cases, we observe a strong positive correlation between predictability and stationarity in the user's browsing, with $r$ equal to $0.6502$ and $0.3320$, respectively. This increase in predictability with the increase in stationarity in users' browsing behavior also complements our results in Section~\ref{sec:impact_design_dec}.
   
\end{itemize}


\subsubsection{Gender-based differences in Browsing Behavior}
While it is interesting that women are more predictable than men, the higher predictability for women in our dataset can potentially be explained by the lower diversity of interests (given by the number of distinct domains visited) for women (M=140.22, $95$\% CI $[132.16, 148.29]$) than men (M=161.46, $95$\% CI $[151.81, 171.13]$), and higher stationarity (captured as mean seconds spent per domain visit) for women (M=37.33, $95$\% CI $[35.54, 39.14]$) than men (M=33.14, $95$\% CI $[31.47, 34.83]$). 

\subsubsection{Age-based differences in Browsing Behavior}
In the analysis presented earlier, we found that the ``middle-aged'' users (35-44) are typically lesser predictable than users of most other ages. We observe that the users in the age group 35-44 have the least amount of stationarity in their browsing with the lowest mean amount of time spent per domain visit (M=29.32, $95$\% CI $[27.21, 31.42]$). This low stationarity could indicate that users in this age group perform a large variety of tasks on the Web. Even among ``young'' users (18-24 and 25-34), higher predictability of users below 24 can partially be explained by 
their higher stationarity (M=34.44, $95$\% CI $[31.17, 37.72]$) as compared to users above 24 (M=29.36, $95$\% CI $[27.74, 30.97]$).

\subsection{Comparing predictability in physical- and cyber-space}
Finally, we briefly compare our computed predictability values for mobility in cyberspace with those presented in prior work for mobility in the physical space. 
We observe that the mean predictability of the next visited location is lower in cyberspace (85\% for stationary and 59\% for non-stationary trajectories) than in physical space (93\% for stationary~\cite{song2010limits} and 71\% for non-stationary~\cite{ikanovic2017alternative} trajectories).\footnote{It is worth noting that while we do compare the predictability results in physical- and cyber-space, it is not clear what the comparable values of $\Delta t$ should be. Since we observe similarly reduced predictability for non-stationary trajectories also, we believe the reduction in predictability in cyberspace is a robust result.} 
Also, while we do observe differences in the distribution of predictability in the cyberspace for users with different demographic attributes, such gender- and age-based differences were not observed for the predictability in the physical space~\cite{song2010limits}.

\section{Discussion \& Future Work}

In this work, we adapted prior work on mobility in physical space to present an information-theoretic framework for estimating the theoretical limits of the predictability of people's mobility on the Web and provided some practical guidelines for its application. We applied the framework to a dataset of German users to demonstrate that people exhibit web routineness. Users' repetitive visitation patterns increase the theoretically achievable predictability of their web browsing behavior to up to 85\%.
We also observed gender- and age-based differences in the predictability of users' web browsing behavior. We highlighted that behavioral differences between demographic groups could partly explain these differences.

\subsubsection{Framework premises}
In our paper, we analyzed the uncertainty embedded in time series (i.e., web trajectories), which enabled us to estimate the predictability upper bounds of users' trajectories on the Web. We highlight two relevant aspects of this framework. First, this upper bound assumes that individuals are somewhat stationary and unlikely to change their web browsing behavior drastically. In our analyses, we failed to recognize drastic changes during the observation period of one month. Future work could examine whether our results hold if one analyzes data that spans a more extended period (e.g., several years). Second, this upper bound tells us the theoretical upper limit to correctly predict an individual's future location  by restricting ourselves to web tracking data only (i.e., location sequences). We show that having just this type of data grants one with a high potential predictive power. However, this upper bound is likely to increase by using different kinds of data, such as users' characteristics (e.g., demographic and behavioral features), visited locations' characteristics (e.g., topical categories), or temporal features (e.g., time of the day when web location was accessed, duration of access); these are also directions for future research.


\subsubsection{Data limitations}
We focused on a dataset limited to one country and desktop browsing only. We are aware that a significant amount of browsing happens over mobile devices. In the future, we would like to compare our results for desktop browsing with those for mobile browsing. 
Moreover, there is always a chance that some of the web browsing activity that we analyze was not generated by humans but by bots. However, we believe this is unlikely for our dataset since the panelists' behavior is closely monitored by the panel company, and the panelists also take part in frequent surveys. In this regard, this type of data is of high quality and allows ethically studying the dynamics of human behavior on the Web.

\subsubsection{Future work}
In the future, we would like to expand our current study to model the impact of more demographic and behavioral attributes of users and to include users from multiple countries to observe cross-cultural differences in the routineness of people on the Web. 
We would also like to explore the definition of distance measures or cost functions of moving from one web location to another to broaden our analysis to use distance-based or cost-aware models of mobility (\eg radius of gyration).
Another line of further research concerns the privacy implications of our work. By progressively removing available browsing history data for a user and studying the change in predictability for the next visited location, we can estimate the privacy risks against an attacker who may get access to a user's partial or full browsing history.
Finally, in this work, we estimate the theoretical limits of predictability achievable on account of the routineness that people demonstrate in their web browsing behavior. Designing specific prediction algorithms to attain this theoretical maximum performance could be pursued in the future.

\bibliography{main}
\bibliographystyle{aaai}

\end{document}